\documentstyle[12pt]{article}
\baselineskip=\normalbaselineskip
\ifx\TwoupWrites\UnDeFiNeD\else\target{\magstepminus1}{11.3in}{8.27in}
	\source{\magstep0}{7.5in}{11.69in}\fi
\newfont{\fourteencp}{cmcsc10 scaled\magstep2}
\newfont{\titlefont}{cmbx10 scaled\magstep2}
\newfont{\authorfont}{cmcsc10 scaled\magstep1}
\newfont{\fourteenmib}{cmmib10 scaled\magstep2}
	\skewchar\fourteenmib='177
\newfont{\elevenmib}{cmmib10 scaled\magstephalf}
	\skewchar\elevenmib='177
\makeatletter
\newcommand\nonsequentialeqnum{
	\@addtoreset{equation}{section}
	\def\theequation{\arabic{section}.\arabic{equation}}}
\newif\ifp@bblock  \p@bblocktrue
\newcommand\nopubblock{\p@bblockfalse}
\newcommand\topspace{\hrule height 0pt depth 0pt \vskip}
\newcommand\p@bblock{\begingroup \tabskip=\hsize minus \hsize
	\baselineskip=1.5\ht\strutbox \topspace-2\baselineskip
	\halign to\hsize{\strut ##\hfil\tabskip=0pt\crcr
	\the\Pubnum\crcr\the\date\crcr}\endgroup}

\renewcommand\titlepage{\ifx\TwoupWrites\UnDeFiNeD\null\vspace{-1.7cm}\fi
	\ifp@bblock\p@bblock \else\hrule height 0pt \relax \fi}
\makeatother
\newtoks\date
\newtoks\Pubnum
\newtoks\pubnum
\Pubnum={UTAP-\the\pubnum}
\date={\today}
\newcommand{\frontpageskip}{\vspace{12pt plus .5fil minus 2pt}}
\renewcommand{\title}[1]{\frontpageskip
	\begin{center}{\titlefont #1}\end{center}\par}
\renewcommand{\author}[1]{\frontpageskip\par\begin{center}
	{\authorfont #1}\end{center}
	\nobreak
	}

\newcommand{\address}[1]{\par\begin{center}{\sl #1}\end{center}\par}
\newcommand{\andaddress}{\par\centerline{\sl and}\address}
\renewcommand{\thanks}[1]{\footnote{#1}}
\renewcommand{\abstract}{\par\frontpageskip\centerline{\fourteencp Abstract}
	\vspace{8pt plus 3pt minus 3pt}}
\newcommand\YITP{\address{Department of Physics \\
	       The University of Tokyo\\
               Tokyo 113,~Japan\\}}

\font\smbf=cmb10 scaled\magstep1
\newcommand{\gtsima}{$\; \buildrel > \over \sim \;$}
\newcommand{\ltsima}{$\; \buildrel < \over \sim \;$}
\newcommand{\simgt}{\lower.5ex\hbox{\gtsima}}
\newcommand{\simlt}{\lower.5ex\hbox{\ltsima}}

\newcommand{\bfx}{{\mbox{\boldmath $x$}}}
\newcommand{\bfy}{{\mbox{\boldmath $y$}}}
\newcommand{\bfz}{{\mbox{\boldmath $z$}}}

\newcommand{\bfk}{{\mbox{\boldmath $k$}}}

\def\pp{\par\parshape 2 0cm 16cm 1cm 15cm\noindent}

\begin{document}    
\thispagestyle{empty}
%
\nonsequentialeqnum 
\pubnum{183/94}
\date{May, 1994}
\titlepage

\title{ANALYTIC EXPRESSION OF THE GENUS IN WEAKLY NON-GAUSSIAN FIELD
INDUCED BY GRAVITY
}

\author{Takahiko Matsubara\thanks{E-mail address: 
matsu@yayoi.phys.s.u-tokyo.ac.jp}
}
\YITP
\andaddress{Department of Physics \\ Hiroshima University \\
Higashi-Hiroshima 724, Japan.}

\baselineskip=18pt
\abstract{
The gravitational evolution of the genus of the density field in
large-scale structure is analytically studied in a weakly nonlinear
regime using second-order perturbation theory. Weakly nonlinear
evolution produces asymmetry in the symmetric genus curve for Gaussian
initial density field. The effect of smoothing the density field in
perturbation theory on the genus curve is also evaluated and gives the
dependence of the asymmetry of the genus curve on spectra of initial
fluctuations.

\medskip

\pp {\it Subject headings}: cosmology: theory --- galaxies: clustering
--- gravitation\\
}

\newpage

\baselineskip=18pt
\setcounter{page}{2}

\begin{center}
\section{\smbf INTRODUCTION}
\end{center}

Recently redshift surveys have been unveiling new detailed information
of structures in the universe. To characterize the pattern of the
structures, the topological analysis of the galaxy distribution
provides a useful and intuitively clear way. Gott, Melott \& Dickinson
(1986) proposed to use the Euler characteristic of surfaces of
constant density as a quantitative measure for the topology of
large-scale structure. The genus, which is widely used in subsequent
topological analyses, is defined by $-1/2$ times the Euler
characteristic per unit volume. More intuitively, the genus
corresponds to ``the number of holes'' of the surfaces minus ``the
number of isolated regions'' surrounded by the surfaces per unit
volume. The genus is a function of smoothing scales and the density
threshold. The genus as a function of density threshold for a fixed
smoothing scale is called the genus curve and is analyzed both in
numerical simulations and in redshift surveys of galaxies by many
authors (Gott, Weinberg \& Melott 1987; Weinberg, Gott \& Melott 1987;
Melott, Weinberg \& Gott 1988; Gott et al. 1989; Park \& Gott 1991;
Park, Gott
\& da Costa 1992; Weinberg \& Cole 1992; Moore et al. 1992; Vogeley,
Park, Geller, Huchra \& Gott 1994; Rhoads, Gott \& Postman 1994).

The only analytical expression for the genus curve known so far is for
Gaussian random density field (Adler 1981; Doroshkevich 1970; Bardeen,
Bond, Kaiser \& Szalay 1986; and Hamilton, Gott \& Weinberg 1986), and
is given by
\begin{equation}
   G(\nu) = \frac{1}{4\pi^2}
	    \left(\frac{\langle k^2 \rangle}{3}\right)^{3/2}   
	    e^{-\nu^2/2} (1-\nu^2),
   \label{eq1}
\end{equation}
where $\nu$ is the difference between density threshold and mean
density in units of standard deviation of density and
\begin{equation}
   \langle k^2 \rangle = 
   \frac{\int k^2 P(k) d^3 k}{\int P(k) d^3 k},
   \label{eq2}
\end{equation}
with $P(k)$ being the power spectrum of the density fluctuation. 
Previous analyses mainly compared the observational genus with the
random Gaussian prediction (\ref{eq1}). With sufficiently large
smoothing scales, this comparison could tell us if initial density
fluctuation is random Gaussian. With the finite smoothing scale of
cosmological interest, the effect of nonlinear gravitational evolution
on the genus curves would be substantial. This nonlinear effect has
been explored only by using $N$-body numerical simulations so far. The
main purpose of this {\it Letter} is to approach analytically this
problem in the weakly nonlinear regime using a second-order
perturbation theory. In the following, the general formula of the
genus curves for the field with weak non-Gaussianity is presented. 
This formula is incorporated in the second order perturbation theory. 
The smoothing effect on perturbation theory is also considered.

\begin{center}
\section{\smbf THE GENUS CURVE FOR QUASI-GAUSSIAN RANDOM FIELD}
\end{center}

Random Gaussian fields are characterized by the fact that connected
correlation functions, except the second-order correlation function
$\xi$, all vanish. This is why the genus curve (eq.[\ref{eq1}]) for
random Gaussian field is completely determined by $P(k)$ which is a
Fourier transform of $\xi$. In this section, correction terms for
equation (\ref{eq1}) in the presence of weak non-Gaussianity are
presented.  The term ``weak non-Gaussianity'' is defined below.

To simplify the notation, the following seven quantities for a
non-Gaussian random field $\delta(x,y,z)$ with zero mean are denoted
by $A_\mu\,(\mu = 1,\ldots,7)$:
\begin{equation}
   \frac{\delta}{\sigma},\quad
   \frac{1}{\sigma} \frac{\partial\delta}{\partial x},\quad
   \frac{1}{\sigma} \frac{\partial\delta}{\partial y},\quad
   \frac{1}{\sigma} \frac{\partial\delta}{\partial z},\quad
   \frac{1}{\sigma} \frac{\partial^2 \delta}{\partial x^2},\quad
   \frac{1}{\sigma} \frac{\partial^2 \delta}{\partial y^2},\quad
   \frac{1}{\sigma} \frac{\partial^2 \delta}{\partial x \partial y},
   \label{eq3}
\end{equation}
where $\sigma \equiv \sqrt{\langle\delta^2\rangle}$ is an {\it rms} of
the field and the field is defined in Cartesian coordinates $x,y,z$.
The field $\delta$ is identified with density contrast
$\rho/\bar{\rho} -1$ in astrophysical applications. The Euler
characteristic per unit volume $n_\chi(\nu)$ of constant surfaces
$\delta = \nu \sigma$ is given by (Adler 1981; Bardeen et al. 1986)
\begin{equation}
   n_\chi(\nu) = \left\langle \rho_\chi(A_\mu) \right\rangle,
   \label{eq4}
\end{equation}
where
\begin{equation}
   \rho_\chi(A_\mu) = 
   \delta_D (A_1 - \nu)
   \delta_D (A_2) \delta_D (A_3) |A_4|
   (A_5 A_6 - A_7^{\,2}),
   \label{eq5}
\end{equation}
and $\delta_D$ is a Dirac's delta-function.  This expression is valid
for general non-Gaussian random fields. Let us proceed to evaluating
the expectation value (eq.[\ref{eq4}]) for weakly non-Gaussian field.
As usual, we define a partition function $Z(J_\mu)$ as a Fourier
transform of a distribution function $P(A_\mu)$ of quantities $A_\mu$:
\begin{equation}
   Z(J_\mu) = \int_{-\infty}^\infty d^7 A P(A_\mu) 
   \exp\left(i \sum_\nu J_\nu A_\nu \right).
   \label{eq6}
\end{equation}
The cumulant expansion theorem (e.g., Ma 1985) states that $\ln
Z(J_\mu)$ is a generating function of connected correlation function
$\psi^{(N)}_{\mu_1\cdots\mu_N} = \langle A_{\mu_1} \cdots A_{\mu_N}
\rangle_{\rm c}$ (see Bertschinger 1992). Then one obtains the inverse
Fourier transform of equation (\ref{eq6}) in the following useful
form:
\begin{equation}
   P(A_\mu) = \exp \left( \sum_{N=3}^{\infty}
   \frac{(-)^N}{N!} \sum_{\mu_1,\ldots,\mu_N}
   \psi^{(N)}_{\mu_1\cdots\mu_N}
   \frac{\partial^N}{\partial A_{\mu_1} \cdots \partial A_{\mu_N}}
   \right) P_{\rm G}(A_\mu),
   \label{eq7}
\end{equation}
where 
\begin{equation}
   P_{\rm G}(A_\mu) = 
   \frac{1}{\sqrt{(2\pi)^7 \det\left(\psi^{(2)}_{\mu\nu}\right)}}
   \exp\left( - \frac{1}{2} \sum_{\mu,\nu} A_\mu 
   \left(\psi^{(2)-1}\right)_{\mu\nu} A_\nu \right),
   \label{eq8}
\end{equation}
is a multivariate Gaussian distribution function characterized by a
correlation matrix $\psi^{(2)}_{\mu\nu}$. In a weakly non-Gaussian
case, the exponential function in equation (\ref{eq7}) is expanded in
Taylor series and equation (\ref{eq4}) is expanded by higher-order
correlations. In the following, we assume that $\psi^{(N)} \sim {\cal
O}(\sigma^{N-2})$. This relation is a very definition of ``weak
non-Gaussianity'' in this letter and is a result of perturbation
theory (Fry 1984; Goroff et al. 1986; Bernardeau 1992). Thus to the
first order in $\sigma$, equation (\ref{eq4}) reduces to
\begin{equation}
   n_\chi(\nu) = 
   \left\langle \rho_\chi (A) \right\rangle_{\rm G}
   + \frac{1}{6} \sum_{\mu,\nu,\lambda} \psi^{(3)}_{\mu\nu\lambda}
     \left\langle
     \frac{\partial^3 \rho_\chi(A)}
          {\partial A_\mu \partial A_\nu \partial A_\lambda}
     \right\rangle_{\rm G}
   + {\cal O}(\sigma^2),
   \label{eq9}
\end{equation}
where $\langle\cdots\rangle_{\rm G}$ denotes averaging by multivariate
Gaussian distribution (eq.[\ref{eq8}]). All the terms in {\it r.h.s.}
of equation (\ref{eq9}) can be evaluated by straightforward but
tedious Gaussian integrals.  Spatial homogeneity and isotropy simplify
the final result as
\begin{equation}
   n_\chi(\nu) = \frac{1}{2\pi^2}
   \left(\frac{\langle k^2 \rangle}{3}\right)^{3/2}
   e^{-\nu^2/2}
   \left[ H_2(\nu)
          + \sigma \left( \frac{S}{6} H_5(\nu)
                          + \frac{3T}{2} H_3(\nu)
                          + 3U H_1(\nu)\right)
          + {\cal O}(\sigma^2)
   \right],
   \label{eq11}
\end{equation}
where $H_n(\nu) = (-)^n e^{\nu^2/2} (d/d\nu)^n e^{-\nu^2/2}$ are
Hermite polynomials, and we have defined three quantities,
\begin{eqnarray}
   && S = \frac{1}{\sigma^4} \langle\delta^3\rangle,
   \nonumber \\
   && T = - \frac{1}{2\langle k^2 \rangle \sigma^4}
            \langle \delta^2 \triangle \delta \rangle,
   \label{eq12} \\
   && U = - \frac{3}{4\langle k^2 \rangle^2 \sigma^4}
            \langle \nabla\delta\cdot\nabla\delta
                    \triangle \delta \rangle.
   \nonumber
\end{eqnarray}
The quantity $S$ is usually called ``skewness''. The first term in
square brackets of equation (\ref{eq11}) corresponds to Gaussian
contribution and the other terms correspond to non-Gaussian
contribution to Euler number density.

As an illustrative application of this result, we consider the case
that correlation functions are given by hierarchical model. In
hierarchical model, connected correlation function of $N$-th order is
modeled as a sum of $N-1$ products of $\xi$, thus our previous
assumption $\psi^{(N)} \sim {\cal O}(\sigma^{N-2})$ is satisfied. 
Specifically, third order correlation function $\zeta(\bfx,\bfy,\bfz)
= \langle \delta(\bfx) \delta(\bfy) \delta(\bfz) \rangle$ is related
to $\xi(|\bfx-\bfy|) = \langle \delta(\bfx) \delta(\bfy) \rangle$ by
\begin{equation}
   \zeta(\bfx,\bfy,\bfz) = Q[\xi(|\bfx-\bfy|)\xi(|\bfy-\bfz|) +
\xi(|\bfy-\bfz|)\xi(|\bfz-\bfx|) + \xi(|\bfz-\bfx|)\xi(|\bfx-\bfy|)],
\label{eq13}
\end{equation}
where $Q$ is a constant (undetermined in this model). If this equation
(\ref{eq13}) is exact for some large smoothing scale such that $\xi(0)
= \sigma^2 \ll 1$, the quantities $S$, $T$ and $U$ reduces to $3Q$,
$2Q$ and $Q$, respectively, and the genus curve with correction terms
of non-Gaussianity is
\begin{equation}
   G^{\rm (hi.)}(\nu) = \frac{1}{4\pi^2}
   \left(\frac{\langle k^2 \rangle}{3}\right)^{3/2}
   e^{-\nu^2/2} \left[ 1 - \nu^2 - \frac{Q}{2}
   \left( \nu^5 - 4 \nu^3 + 3 \nu\right)\sigma \right]
   + {\cal O}(\sigma^2),
   \label{eq14}
\end{equation}
Figure 1(a) plots the results for $Q \sigma = 0, 0.2, 0.4, 0.6$.

\begin{center}
\section{\smbf GRAVITATIONAL EVOLUTION OF THE GENUS CURVE IN SECOND ORDER
PERTURBATION THEORY}
\end{center}

Gravitational nonlinear evolution give rise to $S$, $T$, $U$ even from
the initial Gaussian random density fluctuation which has vanishing
$S$, $T$, $U$. We use second order perturbation theory of the
nonrelativistic collisionless self-gravitating system in the fluid
limit (e.g., Peebles 1980, \S 18) to compute $S$, $T$, $U$ to lowest
order in $\sigma$. Here we present the results explicitly in the
Einstein-de Sitter case, $\Omega = 1$, $\Lambda = 0$. The effects of
parameters $\Omega$, $\Lambda$ on our results are expected to be weak
as discussed in the next section.  Considering growing mode only,
third order correlation function in Fourier space is given by (Fry
1984; Goroff et al. 1986)
\begin{eqnarray}
   && \!\!\!\!
   \left\langle\tilde{\delta}(\bfk_1)
               \tilde{\delta}(\bfk_2)
               \tilde{\delta}(\bfk_3)
   \right\rangle =
   \left\{\left[
      \frac{10}{7} + 
      \left(\frac{k_1}{k_2}+\frac{k_2}{k_1}\right)
         \frac{\bfk_1\cdot\bfk_2}{k_1 k_2} +
      \frac{4}{7}\left(\frac{\bfk_1\cdot\bfk_2}{k_1 k_2}\right)^2
   \right] P(k_1) P(k_2) + {\rm cyc.}\right\} 
   \nonumber \\
   && \qquad\qquad\qquad\qquad\qquad\qquad
   \times (2\pi)^3 \delta_D^3(\bfk_1 + \bfk_2 + \bfk_3),
   \label{eq15}
\end{eqnarray}
where $P(k)$ is a power spectrum of linear theory. Evaluating $S$,
$T$, $U$ (eq.[\ref{eq12}]) in Fourier space with equation (\ref{eq15})
results in,
\begin{equation}
   S = \frac{34}{7},\quad T = \frac{82}{21},\quad U = \frac{54}{35} 
   \label{eq16}
\end{equation}
This value of the skewness $S$ was already given by Peebles (1980). 
Note that $S$, $T$, $U$ are independent on the shape of the initial
power spectrum. The genus curve predicted by perturbation theory is,
therefore,
\begin{equation}
   G^{\rm (p.t.)}(\nu) = \frac{1}{4\pi^2}
   \left(\frac{\langle k^2 \rangle}{3}\right)^{3/2}
   e^{-\nu^2/2} \left[ 1 - \nu^2 - 
   \left( \frac{17}{21} \nu^5 - 
          \frac{47}{21} \nu^3 -
          \frac{4}{5} \nu\right)\sigma \right]
   + {\cal O}(\sigma^2),
   \label{eq17}
\end{equation}
which is plotted in Figure 1(b) for $\sigma = 0, 0.2, 0.4, 0.6$.

In fact, the observable curve is obtained by smoothing of density
fluctuation while the above result is not for smoothed field. Let us
evaluate the smoothing effect in the case that the smoothed density
fluctuation with sufficiently large smoothing scale is well described
by second order perturbation theory. Recently, Juszkiewicz, Bouchet
\& Colombi (1993) gave the smoothing effect on the skewness $S$ for
Gaussian and top-hat filter.  In the following, we use the Gaussian
filter $\delta_R(\bfx) = (2\pi R^2)^{-3/2}\int d^3y \delta(\bfy)
\exp(-|\bfx - \bfy|^2/2R^2)$ which is usually adopted in smoothing noisy
observational data. The quantities $S$, $T$, $U$ for Gaussian smoothed
field $\delta_R$ are obtained similarly as in the unsmoothed case, and
the result is
\begin{eqnarray}
   && S = \frac{3}{28\pi^4 \sigma^4(R)}
   \left[ 5 I_{220} + 7 I_{131} + 2 I_{222}\right],
\nonumber \\
   && T = \left(\frac{\langle k^2 \rangle}{3}\right)^{-1}
   \frac{1}{84\pi^4\sigma^4(R)}
   \left[10 I_{240} + 12 I_{331} + 7 I_{151} + 11 I_{242} +
         2 I_{333}\right],
   \label{eq18} \\
   && U = \left(\frac{\langle k^2 \rangle}{3}\right)^{-2}
   \frac{1}{168\pi^4\sigma^4(R)}
   \left[5 I_{440} + 7 I_{351} - 3 I_{442} - 7 I_{353} -
         2 I_{444}\right],
\nonumber
\end{eqnarray}
where we defined
\begin{equation}
   I_{mnr} = \int_0^\infty dx \int_0^\infty dy \int_{-1}^1 d\mu
   e^{- R^2(x^2 + y^2 + \mu x y)} x^m y^n \mu^r 
   P(x) P(y).
   \label{eq19}
\end{equation}
Wheb $R \rightarrow 0$, the dependence of equation (\ref{eq18}) on the
initial power spectrum is caceled and equation (\ref{eq16}) is
rederived. Thus the genus curve for smoothed field is dependent on the
specific shape of initial power spectrum on the contrary to the
unsmoothed one. For the case of initial power spectrum with power-law,
$P(k) \propto k^n$, equations (\ref{eq18}) are expressed using
generalized hypergeometric functions ${}_pF_q$ as
\begin{eqnarray}
   && S = \frac{30}{7} 
   {}_2F_1\left(\frac{n+3}{2},\frac{n+3}{2};\frac{3}{2};\frac{1}{4}\right)
   - (n+3) 
   {}_2F_1\left(\frac{n+3}{2},\frac{n+5}{2};\frac{5}{2};\frac{1}{4}\right)
   \nonumber \\
   && \qquad   
   + \frac{4}{7}
   {}_3F_2\left(\frac{n+3}{2},\frac{n+3}{2},\frac{3}{2}
                ;\frac{1}{2},\frac{5}{2};\frac{1}{4}\right),
\nonumber \\
   && T = \frac{20}{7}
   {}_2F_1\left(\frac{n+3}{2},\frac{n+5}{2};\frac{3}{2};\frac{1}{4}\right)
   - \frac{4}{7}
   {}_2F_1\left(\frac{n+5}{2},\frac{n+5}{2};\frac{5}{2};\frac{1}{4}\right)
   \nonumber \\
   && \qquad   
   - \frac{n+5}{3}
   {}_2F_1\left(\frac{n+3}{2},\frac{n+7}{2};\frac{5}{2};\frac{1}{4}\right)
   + \frac{22}{21}
   {}_3F_2\left(\frac{n+3}{2},\frac{n+5}{2},\frac{3}{2}
                ;\frac{1}{2},\frac{5}{2};\frac{1}{4}\right)
   \label{eq20} \\
   && \qquad   
   - \frac{2(n+3)}{35}
   {}_3F_2\left(\frac{n+5}{2},\frac{n+5}{2},\frac{5}{2}
                ;\frac{3}{2},\frac{7}{2};\frac{1}{4}\right),
\nonumber \\
   && U = \frac{15}{7}
   {}_2F_1\left(\frac{n+5}{2},\frac{n+5}{2};\frac{3}{2};\frac{1}{4}\right)
   - \frac{n+5}{2}
   {}_2F_1\left(\frac{n+5}{2},\frac{n+7}{2};\frac{5}{2};\frac{1}{4}\right)
   \nonumber \\
   && \qquad   
   -\frac{3}{7}
   {}_3F_2\left(\frac{n+5}{2},\frac{n+5}{2},\frac{3}{2}
                ;\frac{1}{2},\frac{5}{2};\frac{1}{4}\right)
   + \frac{3(n+5)}{10}
   {}_3F_2\left(\frac{n+5}{2},\frac{n+7}{2},\frac{5}{2}
                ;\frac{3}{2},\frac{7}{2};\frac{1}{4}\right)
   \nonumber \\
   && \qquad   
   - \frac{6}{35}
   {}_3F_2\left(\frac{n+5}{2},\frac{n+5}{2},\frac{5}{2}
                ;\frac{1}{2},\frac{7}{2};\frac{1}{4}\right).
\nonumber
\end{eqnarray}
Table 1 shows the numerical values of these quantities for specific
values of $n$.  The corresponding genus curves for $n=1,0,-1,-2$ are
shown in Figure 1(c) to 1(f) for $\sigma = 0, 0.2, 0.4, 0.6$.

\begin{center}
\section{\smbf DISCUSSION}
\end{center}

As seen in equation (\ref{eq11}), the nonlinear correction of first
order in $\sigma$ is an odd function of $\nu$ and generates asymmetry
between high-density region and low-density region in the genus curve,
though it does not change the amplitude $G(0)$. The pattern of the
asymmetry in the genus curve of smoothed density field is dependent on
initial power spectra. Thus, in principle, observations of the genus
curve can restrict the properties of initial fluctuation, such as
Gaussianity, the shape of the spectrum, by the amplitude and the
pattern of asymmetry of the curve. The presently available redshift
data of galaxies are not enough to have the statistically sufficient
accuracy on topology of the large-scale structure. The projects as
Digital Sky Survey (DSS), however, will enable us to have a large
amount of redshift data in near future and the analysis indicated in
this letter will be important one.

We have shown the results explicitly in Einstein-de Sitter case.
Bernardeau (1993) obtained skewness $S$ smoothed by the top-hat filter
for arbitrary $\Omega$ and $\Lambda$ and found that the dependence of
skewness on these parameters is extremely weak. Applying his method,
the Gaussian filtered $S$, $T$ and $U$ can also be evaluated for
arbitrary $\Omega$ and $\Lambda$. Explicit calculations show that the
dependence on these cosmological parameters is also weak and the
Einstein-de Sitter case is a good approximation to the other cases (we
will give the explicit evaluations elsewhere).

The effect of biasing (Kaiser 1984; Bardeen et al. 1986) between the
galaxy distribution and the matter distribution on the genus curve is an
important issue. Quite generally, the correlation functions arose from
local bias approach to the hierarchical model in the large-scale limit
(Szalay 1988; Fry \& Gazta\~naga 1993; Matsubara 1994). The parameter
$Q$ in hierarchical model is determined by biasing mechanism and can
take the positive and negative values. Thus the effect of biasing on
the genus curve is approximately expressed by equation (\ref{eq14}).

The method used in this letter gives the general way to express the
local function of density field by correlation functions: we can
choose any function instead of equation (\ref{eq5}). For example, the
evaluation of nonlinear effect on level crossing statistics (Ryden
1988; Ryden et al. 1988) are straightforward using our method. In the
next work, generalizations to the nonlocal functions of density field
and the multi-point statistics are elegantly described by diagrammatic
language which provides a powerful way to the statistical analyses in
large-scale structure in the universe (Matsubara 1994). Applications
of techniques developed in this letter to the statistics of anisotropy
of cosmic microwave background radiation map are now under progress in
search for non-Gaussianity of primordial fluctuations in the universe.

\vskip1.0cm

I am grateful to thank Y.~Suto for a careful reading of the manuscript
and useful comments.

\newpage

\begin{table}
\begin{center}
\begin{tabular}{cccccccc} \hline\hline
\hfil & unsmoothed & $n=1$ & $n=0$ & $n=-1$ & $n=-2$ & $n=-3$ \\
\hline
$S$ & 4.857 & 3.029 & 3.144 & 3.468 & 4.022 & 4.857 \\ 
$T$ & 3.905 & 2.020 & 2.096 & 2.312 & 2.681 & 3.238 \\ 
$U$ & 1.543 & 1.431 & 1.292 & 1.227 & 1.222 & 1.272 \\
 \hline
\end{tabular}
\end{center}
\caption{The numerical values of $S$, $T$, $U$ for unsmoothed
perturbation theory and smoothed perturbation theory
for power-law spectra, $n=1$ to $-3$. 
\label{tab1}
}
\end{table}

\vspace*{\fill}
\newpage

\baselineskip=15pt
\parskip2pt
\bigskip
\bigskip
\bigskip
\centerline{\bf REFERENCES}
\bigskip

\def\apjpap#1;#2;#3;#4; {\pp#1, {#2}, {#3}, #4}
\def\apjbook#1;#2;#3;#4; {\pp#1, {#2} (#3: #4)}
\def\apjppt#1;#2; {\pp#1, #2.}
\def\apjproc#1;#2;#3;#4;#5;#6; {\pp#1, {#2} #3, (#4: #5), #6}
\apjbook Adler,~R.~J. 1981;
The Geometry of Random Fields;Chichester;Wiley;
\apjpap Bardeen,~J.~M., Bond,~J.~R., Kaiser,~N. \& Szalay,~A.~S. 1986;
ApJ;304;15;
\apjpap Bernardeau,~F. 1992;ApJ;392;1;
\apjppt Bernardeau,~F. 1993;CITA 93/44 preprint;
\apjbook Bertschinger, E. 1992;in {\it New Insights into the
   Universe}, eds. Mart\'{\i}nez, V.~J., Portilla,~M. \& S\'aez,~D.;
   Springer-Verlag;Berlin Heidelberg;
\apjpap Doroshkevich,~A.~G. 1970;Astrophysics;6;320 (transl. from
Astrofizika, 6, 581);
\apjpap Fry,~J.~N. 1984;ApJ;279;499;
\apjpap Fry,~J.~N. \& Gazta\~naga,~E. 1993;ApJ;413;447;
\apjpap Goroff,~M.~H., Grinstein,~B., Rey,~S.-J. \& Wise,~M.~B. 1986;ApJ;311;6;
\apjpap Gott,~J.~R., Miller,~J., Thuan,~T.~X., Schneider,~S.~E.,
Weinberg,~D.~H., Gammie,~C., Polk,~K., Vogeley,~M., Jeffrey,~S.,
Bhavsar,~S.~P., Melott,~A.~L., Giovanelli,~R., Haynes,~M.~P.,
Tully,~R.~B. \& Hamilton,~A.~J.~S. 1989;ApJ;340;625; 
\apjpap Gott,~J.~R., Melott,~A.~L. \& Dickinson,~M. 1986;ApJ;306;341;
\apjpap Gott,~J.~R., Weinberg,~D.~H. \& Melott,~A.~L. 1987;ApJ;319;1;
\apjpap Hamilton,~A.~J.~S., Gott,~J.~R. \& Weinberg,~D. 1986;
ApJ;309;1;
\apjpap Juszkiewicz,~R., Bouchet,~F.~R. \& Colombi,~S. 1993;ApJ;412;L9;
\apjpap Kaiser, N. 1984;ApJ;284;L9;
\apjbook Ma,~S.-K. 1985;
Statistical Mechanics;Philadelphia;World Scientific;
\apjppt Matsubara,~T. 1994;in preparation;
\apjpap Melott,~A.~L., Weinberg,~D.~H. \& Gott,~J.~R. 1988;ApJ;328;50;
\apjpap Moore,~B., Frenk,~C.~S., Weinberg,~D.~H., Saunders,~W.,
Lawrence,~A., Ellis,~R.~S., Kaiser,~N., Efstathiou,~G. \&
Rowan-Robinson,~M. 1992;MNRAS;256;477;
\apjpap Park,~C., Gott,~J.~R. \& da Costa,~L.~N. 1992;ApJ;392;L51;
\apjpap Park,~C. \& Gott,~J.~R. 1991;ApJ;378;457;
\apjbook Peebles,~P.~J.~E. 1980;The Large-Scale Structure of the
Universe;Princeton University Press;Princeton;
\apjpap Rhoads,~J.~E., Gott,~J.~R. \& Postman,~M. 1994;ApJ;421;1;
\apjpap Ryden,~B.~S. 1988;ApJ;333;L41;
\apjpap Ryden,~B.~S., Melott,~A.~L., Craig,~D.~A., Gott,~J.~R.,
Weinberg,~D.~H., Scherrer,~R.~J., Bhavsar,~S.~P. \& Miller,~J.~M.
1988;ApJ;340;647; 
\apjpap Szalay,~A.~S. 1988;ApJ;333;21;
\apjpap Vogeley,~M.~S., Park.~C., Geller,~M.~J., Huchra,~J.~P. \&
Gott,~J.~R. 1994;ApJ;420;525;
\apjpap Weinberg,~D.~H. \& Cole,~S. 1992;MNRAS;259;652;
\apjpap Weinberg,~D.~H., Gott,~J.~R. \& Melott,~A.~L. 1987;ApJ;321;2;

\vfill\eject

\bigskip
\bigskip
\baselineskip=18pt

\centerline{\bf FIGURE CAPTIONS}

\bigskip

\pp {\bf Figure 1 :} Asymmetries of the normalized genus curves
$G(\nu)/G(0)$ induced by non-Gaussianity to first order of {\it rms}
$\sigma$ of fluctuation. The sources of non-Gaussianity are (a)
hierarchical model, (b) unsmoothed perturbation theory with Gaussian
initial fluctuations, Gaussian-smoothed perturbation theory with
Gaussian initial fluctuations of power-law spectra, (c) $P(k) \propto
k$, (d) $P(k) = {\rm const.}$, (e) $P(k) \propto k^{-1}$, (f) $P(k)
\propto k^{-2}$. Solid lines, dotted lines, dashed lines, long-dashed
lines show $\sigma =$ (in (a), $Q\sigma =$) $0, 0.2, 0.4, 0.6$ cases,
respectively.

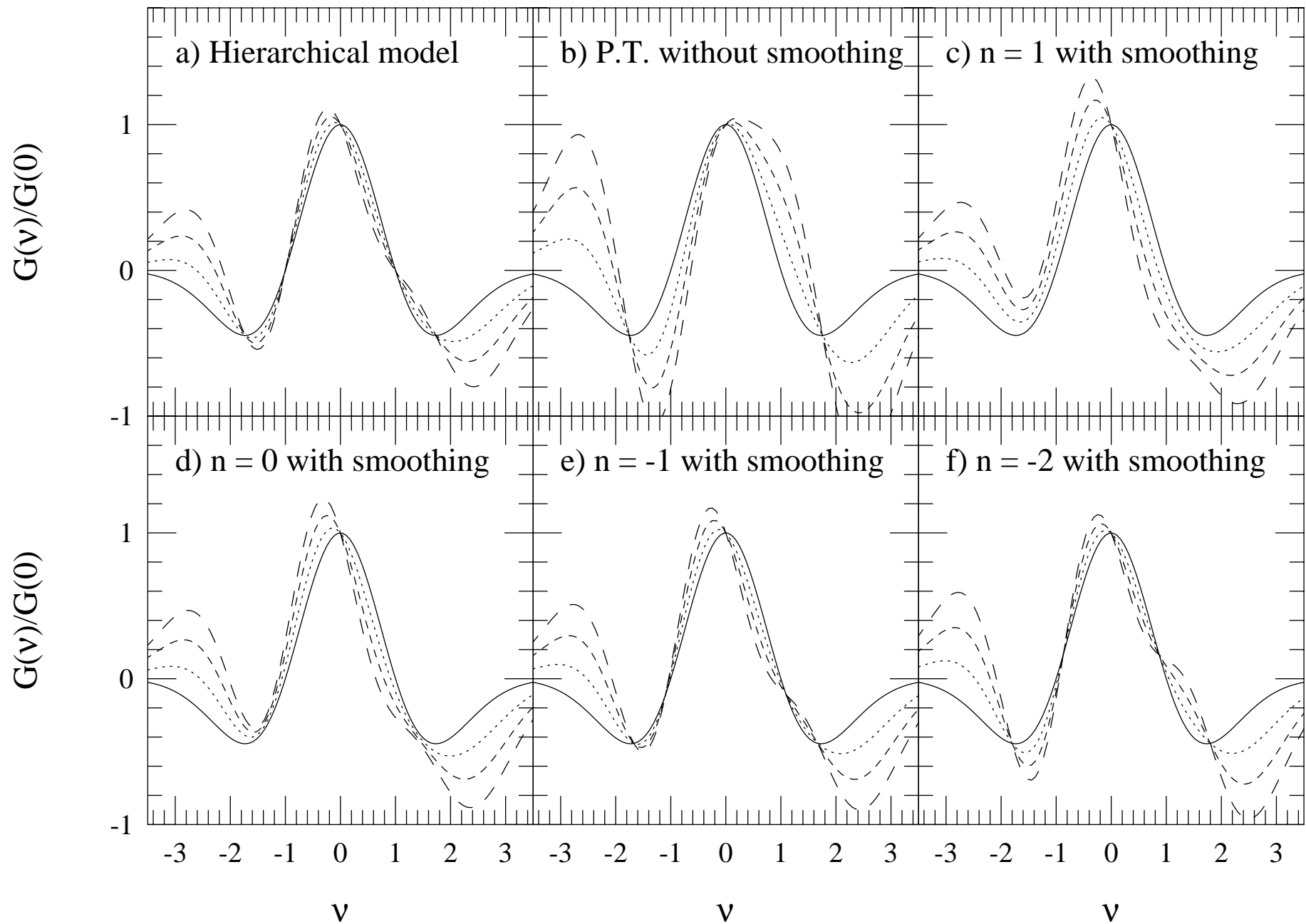
\end{document}